\documentclass[english,12pt]{article}
\usepackage[cp1251]{inputenc}
\usepackage{babel}
\usepackage{amsmath, amssymb}
\usepackage{axodraw}
\usepackage{epsfig}
\begin{document}

\title{\bf Production of two electron-positron couples in electroweak $\gamma\gamma$- interaction} \vspace{0.5cm}
\author{T. Shishkina, I. Sotsky. \\  \it  NC PHEP, BSU, Minsk, Belarus}
\date{}
\maketitle
\begin{abstract}
{\small We present the calculation of  $\gamma\gamma \rightarrow
2e^+e^-$ process cross section. The construction are perfomed
using both helicity amplitude method and method of precision
covariant calculation. The magnitude of cross section is obtained
by the Monte-Carlo method of numerical integration. Different
energies, polarization states and kinematics cuts are considered.}
\end{abstract}

\section{Introduction}

 At future linear collider besides of  $e^-e^-$ and $e^+e^-$
interactions    $\gamma\gamma$ and $\gamma\,e$ modes  are planned
to realize (TESLA, CLIC and others). This  possibility will
provide a great advantage in study of non-Abelian nature of
electroweak interaction, gauge boson coupling as well as couplings
of gauge bosons with Higgs particles if it is light enough to be
produced. Since $W^{\pm}$ and Higgs bosons decay within detector
they can be investigated via their decay products, for instance
four leptons in final state. Because of high accuracy and
relatively clean environment provided by a leptonic collider, a
precision calculation of backgrounds of $\gamma\gamma
\rightarrow...\rightarrow 4l$ processes is necessary.

Total cross sections of processes $\gamma\gamma \rightarrow
2e^-2e^+,$ $\gamma\gamma \rightarrow e^+e^-\mu^+\mu^-,$
$\gamma\gamma \rightarrow 2\mu^-2\mu^+$  have been already
calculated \cite{c1}-\cite{c3} about 30 years ago and were found
to be large. However there was used the low energy approximation,
and obtained results are not applicable to analyze the results of
high energy experiments.

The matrix element of $\gamma\gamma\;\rightarrow 4l$ process has
been constructed also in ref. \cite{c4}. However at that paper
neither calculation of cross section no numerical analyze are
present. So it is impossible to perform any numerical congruence.

This process was also analysed in ref. \cite{c5} where some
numerical calculations  were performed, and the dependence of
total cross section from the energy of initial beam was
investigated. But authors consider low energy region only (about
1-5 GeV) while high energy experiments require to study
$\gamma\gamma$- interaction at beam energy  up to 300-500 GeV.

Beside that the process of four lepton production in
$\gamma\,\gamma$ interaction  was considered in ref. \cite{c6}.
There was applied the algorithm ALPHA for automatic computations
of scattering amplitude. However modern high energy experiments
require the calculation of cross section at definite polarization
states of initial and final particles that ALPHA method doesn't
provide.

\section{Construction and calculations}

There are six topologically different Feynman diagrams of
electroweak interaction describing process
$\gamma\gamma\rightarrow 4l$ (see fig.1).  Whole set of diagrams
can be derived on base of these six ones  using C- P- and crossing
symmetries.

$$ $$
 \vspace{2cm}
\begin{center}
\begin{picture}(400,100)
\Photon(10,180)(35,167){1}{2} \Photon(10,100)(35,114){1}{2}
\ArrowLine(35,167)(85,175) \ArrowLine(85,106)(35,114)
\ArrowLine(45,150)(35,167) \ArrowLine(35,114)(45,130)
\Photon(45,150)(45,130){1}{2} \ArrowLine(85,150)(45,150)
\ArrowLine(45,130)(85,130) \Text(15,185)[l]{\small $\gamma(k_1)$}
\Text(2,115)[l]{\small $\gamma(k_2)$}\Text(90,177)[l]{\small
$p_1$}\Text(90,150)[l]{\small $p_2$} \Text(90,130)[l]{\small
$p_3$}\Text(90,104)[l]{\small $p_4$} \Text(35,95)[l]{\small
$(1)$}\Text(7,141)[l]{\small
$W,Z,\gamma$}\Vertex(35,167){2}\Vertex(35,114){2}\Vertex(45,150){2}
\Vertex(45,130){2}

\Photon(150,180)(175,167){1}{2} \Photon(150,100)(175,114){1}{2}
\ArrowLine(175,167)(225,175) \ArrowLine(225,106)(175,114)
\ArrowLine(185,140)(175,167) \ArrowLine(175,114)(185,140)
\Photon(185,140)(205,140){1}{2} \ArrowLine(225,130)(205,140)
\ArrowLine(205,140)(225,150) \Text(155,185)[l]{\small
$\gamma(k_1)$} \Text(150,117)[l]{\small
$\gamma(k_2)$}\Text(230,177)[l]{\small
$p_1$}\Text(230,150)[l]{\small $p_3$} \Text(230,130)[l]{\small
$p_4$}\Text(230,104)[l]{\small $p_2$} \Text(175,95)[l]{\small
$(2)$}\Text(184,153)[l]{\small
$W,Z,\gamma$}\Vertex(175,167){2}\Vertex(175,114){2}\Vertex(205,140){2}
\Vertex(185,140){2}

\Photon(290,180)(315,167){1}{2} \Photon(290,100)(315,114){1}{2}
\ArrowLine(315,167)(345,159) \ArrowLine(365,106)(315,114)
\ArrowLine(315,114)(315,167) \ArrowLine(345,159)(365,143)
\Photon(345,159)(355,169){1}{2} \ArrowLine(365,158)(355,169)
\ArrowLine(355,169)(365,180) \Text(290,185)[l]{\small
$\gamma(k_1)$} \Text(290,117)[l]{\small
$\gamma(k_2)$}\Text(370,185)[l]{\small
$p_1$}\Text(370,160)[l]{\small $p_2$} \Text(370,145)[l]{\small
$p_3$}\Text(370,104)[l]{\small $p_4$} \Text(315,95)[l]{\small
$(3)$}\Text(317,172)[l]{\small
$W,Z,\gamma$}\Vertex(315,167){2}\Vertex(315,114){2}\Vertex(345,159){2}
\Vertex(355,169){2}

\end{picture}
\end{center}

\vspace{0cm}

\begin{center}
\begin{picture}(400,100)
\Photon(10,180)(50,140){1}{4} \Photon(10,100)(50,140){1}{4}
\ArrowLine(65,155)(85,175) \ArrowLine(85,106)(65,125)
\Photon(50,140)(65,155){1}{2}\Photon(50,140)(65,125){1}{2}
\ArrowLine(85,150)(65,155) \ArrowLine(65,125)(85,130)
\Text(15,185)[l]{\small $\gamma(k_1)$} \Text(-3,117)[l]{\small
$\gamma(k_2)$}\Text(90,177)[l]{\small
$p_1$}\Text(90,150)[l]{\small $p_2$} \Text(90,130)[l]{\small
$p_3$}\Text(90,104)[l]{\small $p_4$} \Text(35,95)[l]{\small
$(4)$}\Text(48,155)[l]{\small $W$}\Text(48,125)[l]{\small
$W$}\Vertex(50,140){2}\Vertex(65,155){2}\Vertex(65,125){2}

\Photon(150,180)(175,160){1}{3} \Photon(150,100)(175,120){1}{3}
\ArrowLine(200,160)(225,175) \ArrowLine(225,106)(200,120)
\Photon(175,120)(175,160){1}{3}
\Photon(175,120)(200,120){1}{2}\Photon(175,160)(200,160){1}{2}
\ArrowLine(200,120)(225,130) \ArrowLine(225,150)(200,160)
\Text(155,185)[l]{\small $\gamma(k_1)$} \Text(135,114)[l]{\small
$\gamma(k_2)$}\Text(230,177)[l]{\small
$p_1$}\Text(230,150)[l]{\small $p_2$} \Text(230,130)[l]{\small
$p_3$}\Text(230,104)[l]{\small $p_4$} \Text(175,95)[l]{\small
$(5)$}\Text(183,168)[l]{\small $W$}\Text(183,130)[l]{\small
$W$}\Text(183,168)[l]{\small $W$} \Text(160,142)[l]{\small $W$}
\Vertex(175,160){2}\Vertex(175,120){2}\Vertex(200,120){2}
\Vertex(200,160){2}

\Photon(290,180)(315,167){1}{2} \Photon(290,100)(315,114){1}{2}
\Photon(315,167)(350,167){1}{3} \ArrowLine(365,106)(315,114)
\ArrowLine(315,114)(335,134) \ArrowLine(335,134)(365,134)
\Photon(315,167)(335,134){1}{3} \ArrowLine(365,158)(350,167)
\ArrowLine(350,167)(365,180) \Text(295,185)[l]{\small
$\gamma(k_1)$} \Text(279,110)[l]{\small
$\gamma(k_2)$}\Text(370,185)[l]{\small
$p_1$}\Text(370,160)[l]{\small $p_2$} \Text(370,135)[l]{\small
$p_3$}\Text(370,104)[l]{\small $p_4$} \Text(315,95)[l]{\small
$(6)$}\Text(332,175)[l]{\small $W$}\Text(310,145)[l]{\small
$W$}\Vertex(315,167){2}\Vertex(315,114){2}\Vertex(350,167){2}
\Vertex(335,134){2}

\end{picture}
\end{center}
\vspace{-3.8cm}
\begin{center}
{\small {$Fig. \;1.$} Feynman diagrams for process $\gamma\gamma
\rightarrow 4l$.}
\end{center}

The diagrams  containing charged current exchange are excluded
because only processes with four charged leptons in final state
are considered. Matrix elements for  remaining diagrams (1)-(3)
have the following form:  \vspace{0.5cm}

\begin{eqnarray}\label{c1}
\begin{array}{c}
 M_1 =
\frac{{\displaystyle-ie^4}}{\displaystyle(k_1-p_1-p_2)^2}\overline{u}(p_1)\widehat{\varepsilon}(k_1)\frac{\displaystyle\widehat{p}_1-\widehat{k}_1+m}{\displaystyle(p_1-k_1)^2-m^2}\gamma^{\mu}v(p_2)\overline{u}(p_3)\gamma_{\mu}\times
\\ \frac{\displaystyle\widehat{k}_2-\widehat{p}_4+m}{\displaystyle(k_2-p_4)^2-m^2}
  \widehat{\varepsilon}(k_2)v(p_4)-

{ie^2\left(\frac{\displaystyle g}{\displaystyle
2cos(\theta_W)}\right)^2}{D_{\mu\nu}(k_1-p_1-p_2)}\overline{u}(p_1)

\widehat{\varepsilon}(k_1)\times  \\
\frac{\displaystyle\widehat{p}_1-\widehat{k}_1+m}{\displaystyle(p_1-k_1)^2-m^2}\gamma^{\mu}
 (g_V+g_A\gamma_5)v(p_2)

\overline{u}(p_3)\gamma^{\nu}(g_V+g_A\gamma_5)\frac{\displaystyle\widehat{k}_2-\widehat{p}_4+m}{\displaystyle(k_2-p_4)^2-m^2}\widehat{\varepsilon}(k_2)v(p_4),\!\!\!\!\!\!\!\!\!\!\!\!

\end{array}
\end{eqnarray}

\begin{eqnarray}\label{c2}
\begin{array}{c}

M_2 =
\frac{{\displaystyle-ie^4}}{\displaystyle(p_3+p_4)^2}\overline{u}(p_1)\widehat{\varepsilon}(k_1)\frac{\displaystyle\widehat{p}_1-\widehat{k}_1+m}{\displaystyle(p_1-k_1)^2-m^2}\gamma^{\mu}\frac{\displaystyle\widehat{k}_2-\widehat{p}_2+m}{\displaystyle(k_2-p_2)^2-m^2}\widehat{\varepsilon}(k_2)v(p_2)
\\ \times \overline{u}(p_3)\gamma_{\mu}v(p_4)-
{ie^2\left(\frac{\displaystyle g}{\displaystyle
2cos(\theta_W)}\right)^2}{D_{\mu\nu}(p_3+p_4)}\overline{u}(p_1)\widehat{\varepsilon}(k_1)\frac{\displaystyle\widehat{p}_1-\widehat{k}_1+m}
{\displaystyle(p_1-k_1)^2-m^2}\gamma^{\mu}\times
\\ (g_V+g_A\gamma_5)
\frac{\displaystyle\widehat{k}_2-\widehat{p}_2+m}{\displaystyle(k_2-p_2)^2-m^2}\widehat{\varepsilon}(k_2)v(p_2)\overline{u}(p_3)\gamma^{\nu}(g_V+g_A\gamma_5)v(p_4),
\end{array}
\end{eqnarray}

\begin{eqnarray}\label{c3}
\begin{array}{c}

 M_3
=\frac{{\displaystyle-ie^4}}{\displaystyle(p_1+p_2)^2}\overline{u}(p_3)\gamma^{\mu}\frac{\displaystyle\widehat{p}_1+\widehat{p}_2+\widehat{p}_3+m}{\displaystyle(p_1+p_2+p_3)^2-m^2}\widehat{\varepsilon}(k_1)\frac{\displaystyle\widehat{k}_2-\widehat{p}_4+m}{\displaystyle(k_2-p_4)^2-m^2}\times
\\ \widehat{\varepsilon}(k_2)v(p_4)
\overline{u}(p_1)\gamma_{\mu}v(p_2)-
 {ie^2\left(\frac{\displaystyle g}{\displaystyle
2cos(\theta_W)}\right)^2}{D_{\mu\nu}(p_1+p_2)}\overline{u}(p_3)\gamma^{\mu}(g_V+g_A\gamma_5)\times
\\
\frac{\displaystyle\widehat{p}_1+\widehat{p}_2+\widehat{p}_3+m}{\displaystyle(p_1+p_2+p_3)^2-m^2}
 \widehat{\varepsilon}(k_1)

\frac{\displaystyle\widehat{k}_2-\widehat{p}_4+m}{\displaystyle(k_2-p_4)^2-m^2}\widehat{\varepsilon}(k_2)v(p_4)\overline{u}(p_1)\gamma^{\nu}(g_V+g_A\gamma_5)v(p_2).
\end{array}
\end{eqnarray}

\vspace{0.5cm} \noindent  Here $\widehat{p}_1 =
p_1^{\mu}\gamma_{\mu}$, where $p_1^{\mu}$ is $\mu$ - component of
four momentum $p_1$;
$\widehat{\varepsilon}(k_1)=\varepsilon^{\mu}(k_1)\gamma_{\mu}$,
where $\varepsilon^{\mu}(k_1)$ --  $\mu$ - component of
polarization vector of photon with four momentum $k_1$,
$D_{\mu\nu}(q)$ --  propagator of $Z^0$- boson with momentum $q$.

   Corresponding cross section has the
form:
\begin{eqnarray}\label{c2}
\begin{array}{c}

\sigma = \frac{\displaystyle 1}{\displaystyle
4(k_1k_2)}\int|M|^2d\Gamma, \large
\end{array}
\end{eqnarray}

\noindent where $$d\Gamma
=
\frac{d^3p_1}{(2\pi)^32p_1^0}\frac{d^3p_2}{(2\pi)^32p_2^0}\frac{d^3p_3}{(2\pi)^32p_3^0}\frac{d^3p_4}{(2\pi)^32p_4^0}(2\pi)^4\delta(k_1+k_2-p_1-p_2-p_3-p_4)$$

\noindent is phase space element.

\par
 In the present paper  squared  matrix elements  are constructed  using both helicity
 amplitude method \cite{c7}-\cite{c10} and  method of precision
covariant calculation (see, for example, refs.
\cite{c11},\cite{c12}). The helicity amplitude method allows to
calculate  cross section directly for each definite polarization
state of initial and final particles. The matrix element
constructed by this method consists of invariants without any
bispinor, so many difficulties in squaring and numerical
integration are excluded. The explicit form of all amplitudes
obtained in frame of helicity amplitude method one can find in
ref. \cite{c13}.
   The method of covariant calculation allows to
obtain the matrix element without any approximation and was used
for verification of  results in each  step of our construction and
calculation.

 For the investigation of total and differential cross section
 the Monte-Carlo method of numerical integration    was applied.
 If two or more produced particles propagate   very closely, the square of
matrix element becomes very large. (So-called collinear peak
problem is arisen.) To achieve the acceptable precision the method
of Monte-Carlo was adopted. Instead of  regular distribution of
kinematic variables (such distribution is usually applied in
Monte-Carlo generators) we have used irregular one, which is very
closed  to matrix element behavior. This proximity can be obtained
by choosing of several free parameters available in the
distribution function. So the adopted Monte-Carlo generator gives
the results with very  small numerical error (about $0.5\% -
0.7\%$).

The accuracy of approach based on the helicity amplitude method
was estimated by comparing with  the precision covariant one.
Since the results of both method for cross section of two
electron-positron pair production have excellent agreement at each
kinematics point, the mass contribution is practically negligible
at least if TESLA energy and cuts are used.

\vspace{0.5cm}

\section{Conclusion}

In this paper the squared matrix elements of process $\gamma\gamma
\rightarrow 2e^-2e^+$ have been constructed using   of the
helicity amplitude method as well as the method of precision
covariant calculations. Numerical integration of obtained cross
sections were performed using adopted Monte-Carlo generator. The
value of differential and total cross section both at averaged and
fixed polarization states were calculated at different energy and
kinematics cuts on polar angles.

\begin{figure}[h!]
\leavevmode
\begin{minipage}[b]{.5\linewidth}
\centering
\includegraphics[width=\linewidth, height=6.5cm, angle=0]{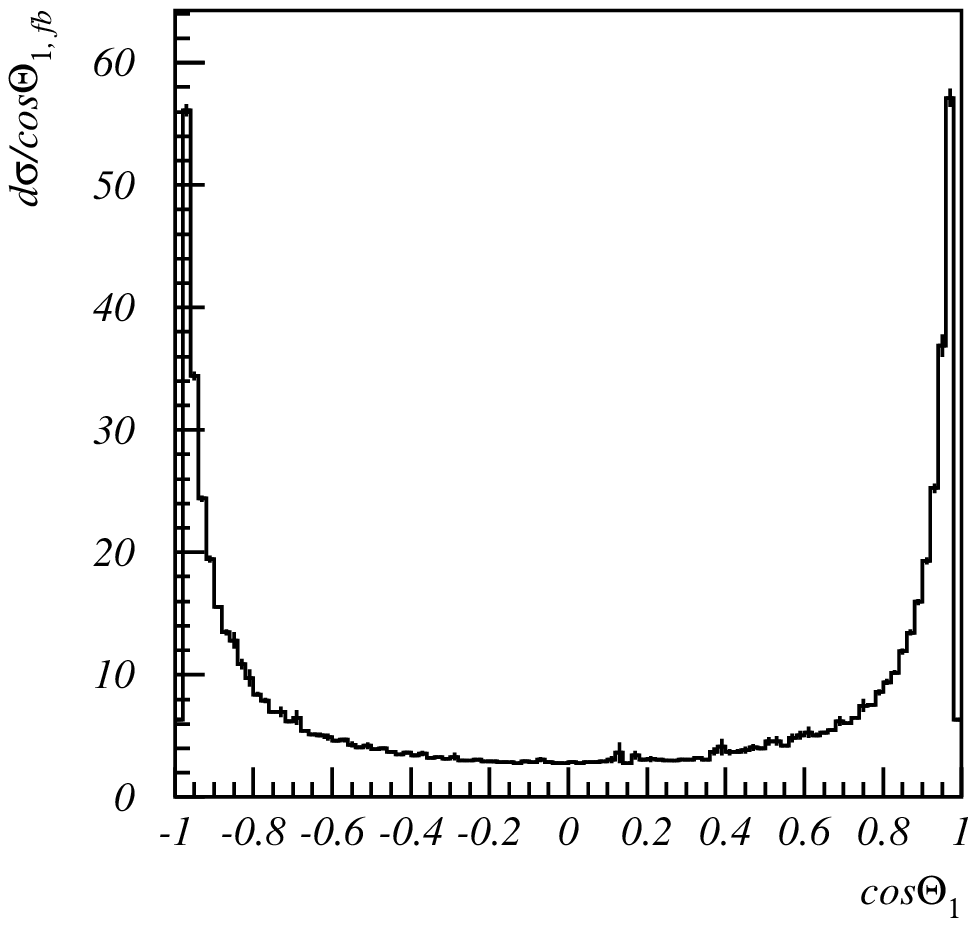}
\end{minipage}\hfill
\begin{minipage}[b]{.5\linewidth}
\centering
\includegraphics[width=\linewidth, height=6.5cm, angle=0]{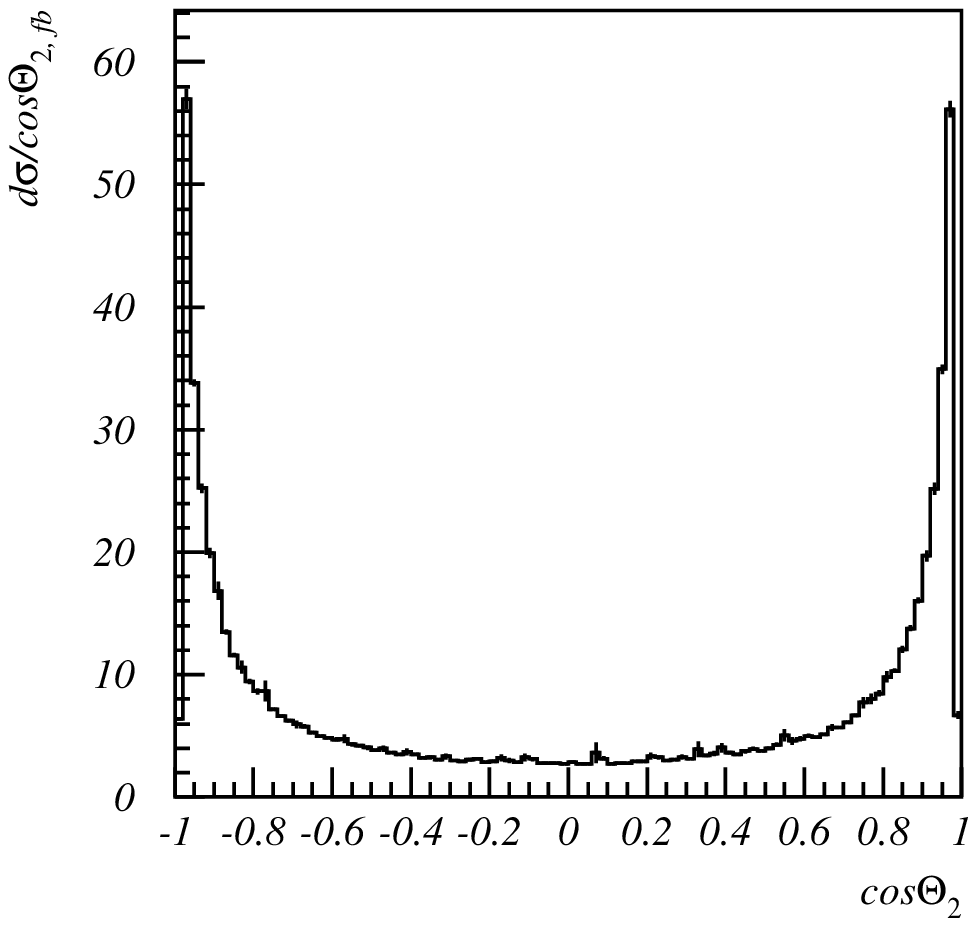}
\end{minipage}

\vspace{-10pt}
{\small Fig.2. Spin average differential cross section of
$\gamma\gamma \rightarrow 2e^-2e^+$ process at c.m. energy of
$\gamma\gamma-$ beam 0.5 TeV. $\theta_{1(2)}$ is the angle between
the directions of the first(second)  photon and the electron.
 The values of polar angle cut and cut of
angle between any final particles are $11^o$ and $3^o$
respectively.}

\end{figure}

\newpage
\begin{center}   Table.1. The dependence of total cross
section on energies and  kinematics cuts. Here the notation
$(\alpha,\beta)$ for describing of kinematics cuts is used, where
$\alpha$ is the cut of angle between directions of any  final
particles, $\beta$ -- the cut of polar angle.

\end{center} \vspace{-0.2cm}
\begin{center}
\begin{tabular}{|c|c|c|c|c|c|c|}

 \hline
  energy (TeV)& cut & $\sigma$ (fb) \\   \hline
$0.3\vphantom{\Bigr[}$&$(3^o,7^o)$&$76.41\pm 0.47$ \\ \hline
$0.3\vphantom{\Bigr[}$&$(3^o,11^o)$&$35.38\pm 0.23$ \\ \hline
$0.5\vphantom{\Bigr[}$&$(3^o,7^o)$&$31.96\pm 0.19$ \\ \hline
$0.5\vphantom{\Bigr[}$&$(3^o,11^o)$&$15.32\pm 0.11$ \\ \hline
$1\vphantom{\Bigr[}$&$(3^o,7^o)$&$9.90\pm 0.07$ \\ \hline
$1\vphantom{\Bigr[}$&$(3^o,11^o)$&$4.81\pm 0.03$ \\ \hline

\end{tabular}
\end{center}
\vspace{1cm}

\vspace{1cm}

 {\normalsize  Following notation is used in fig. 3 and
fig. 4 for describing spin configuration: $(+,-,-,+,-,+)$ means
$(\lambda_1=+1,\lambda_2=-1,\lambda_3=-1,\lambda_4=+1,\lambda_5=-1,\lambda_6=+1)$;
$(+,+,+,+,-,-)$ means
$(\lambda_1=+1,\lambda_2=+1,\lambda_3=+1,\lambda_4=+1,\lambda_5=-1,\lambda_6=-1)$,
where  $\lambda_{1,2}\;$ corresponds to polarization of photon
with four momentum $k_{1,2}$;  $\lambda_{3,4,5,6}\;-$ helicity of
lepton with four momentum $p_{1,2,3,4}$\;respectively}.

\vspace{-0.7cm}
\begin{figure}[h!]
\leavevmode
\begin{minipage}[b]{.5\linewidth}
\centering
\includegraphics[width=\linewidth, height=6.5cm, angle=0]{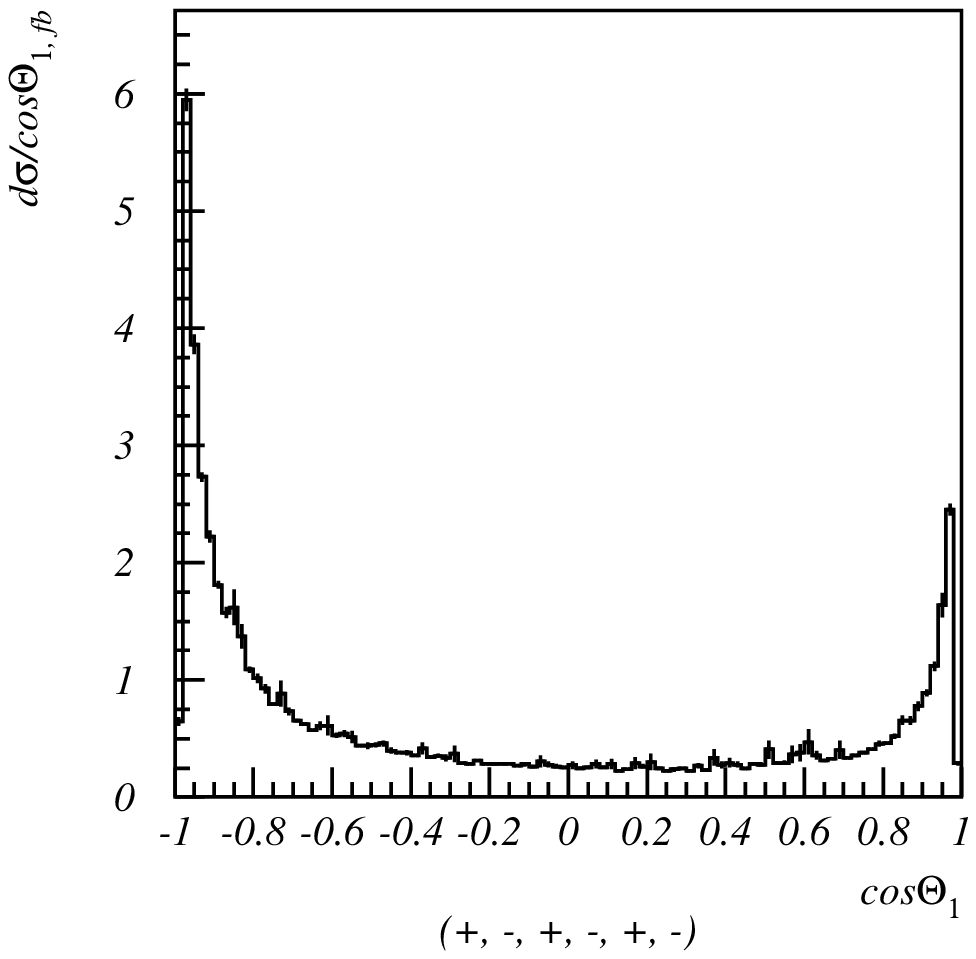}
\end{minipage}\hfill
\begin{minipage}[b]{.5\linewidth}
\centering
\includegraphics[width=\linewidth, height=6.5cm, angle=0]{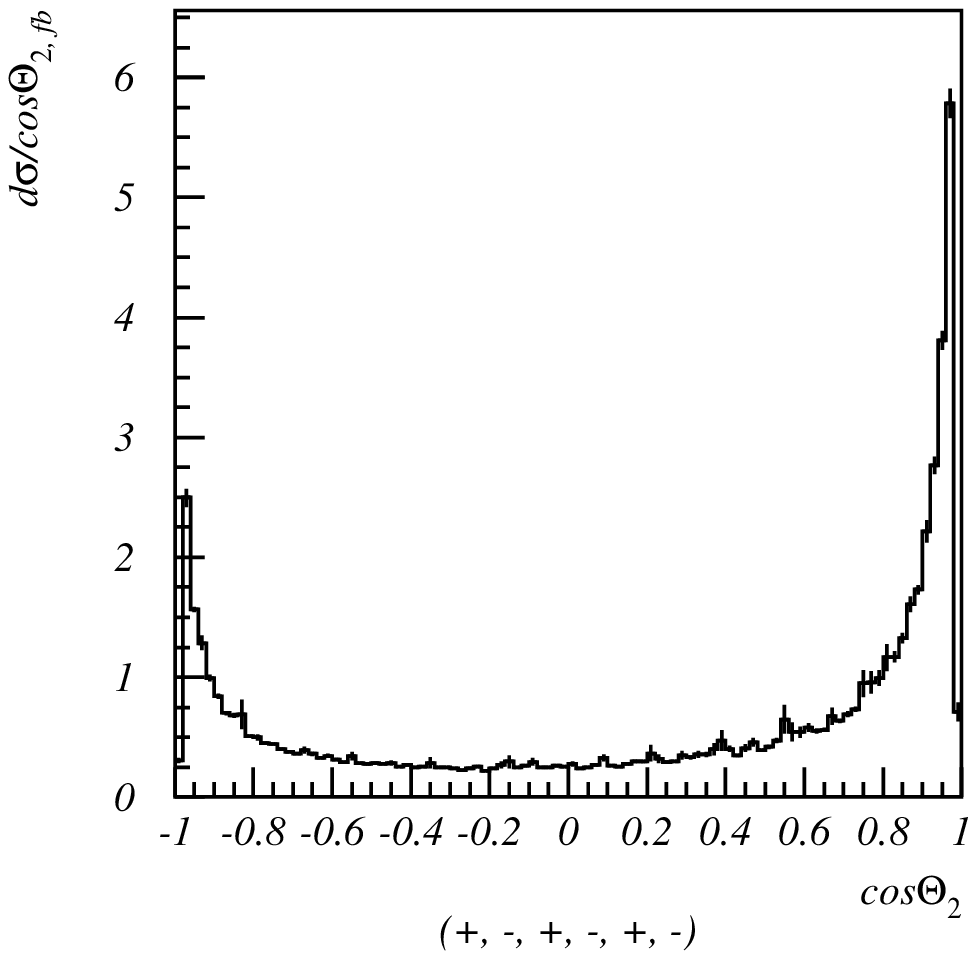}
\end{minipage}

\vspace{-10pt}
{\small Fig.3. The differential cross section of $\gamma\gamma
\rightarrow 2e^-2e^+$ process at c.m. energy of $\gamma\gamma-$
beam  $ 0.5\;{\rm TeV}$  at fixed polarization states of
interacting particles. $\theta_{1(2)}$ is the angle between the
directions of the first(second) photon and the  electron.  The
values of polar angle cut and cut of angle between any final
particles are $11^o$ and $3^o$ respectively.}
\end{figure}
\vspace{0.7cm}

\begin{figure}[h!]
\leavevmode
\begin{minipage}[b]{.5\linewidth}
\centering
\includegraphics[width=\linewidth, height=6.5cm, angle=0]{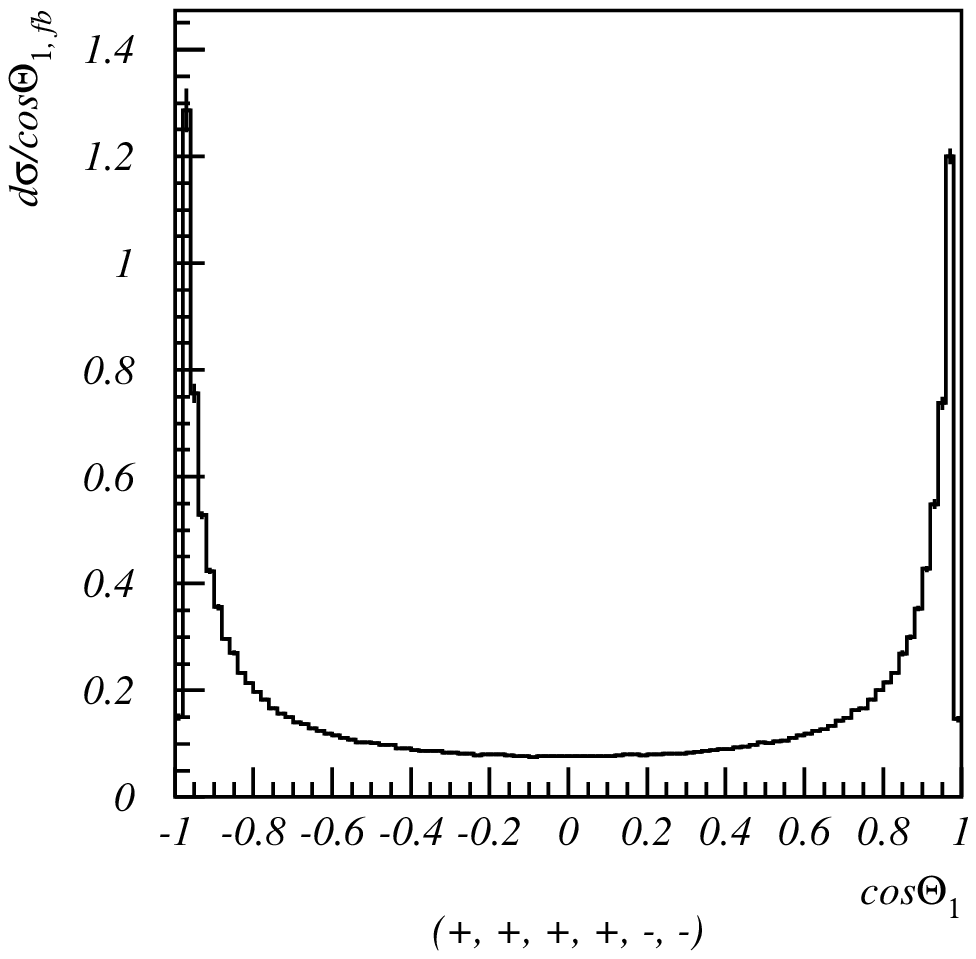}
\end{minipage}\hfill
\begin{minipage}[b]{.5\linewidth}
\centering
\includegraphics[width=\linewidth, height=6.5cm, angle=0]{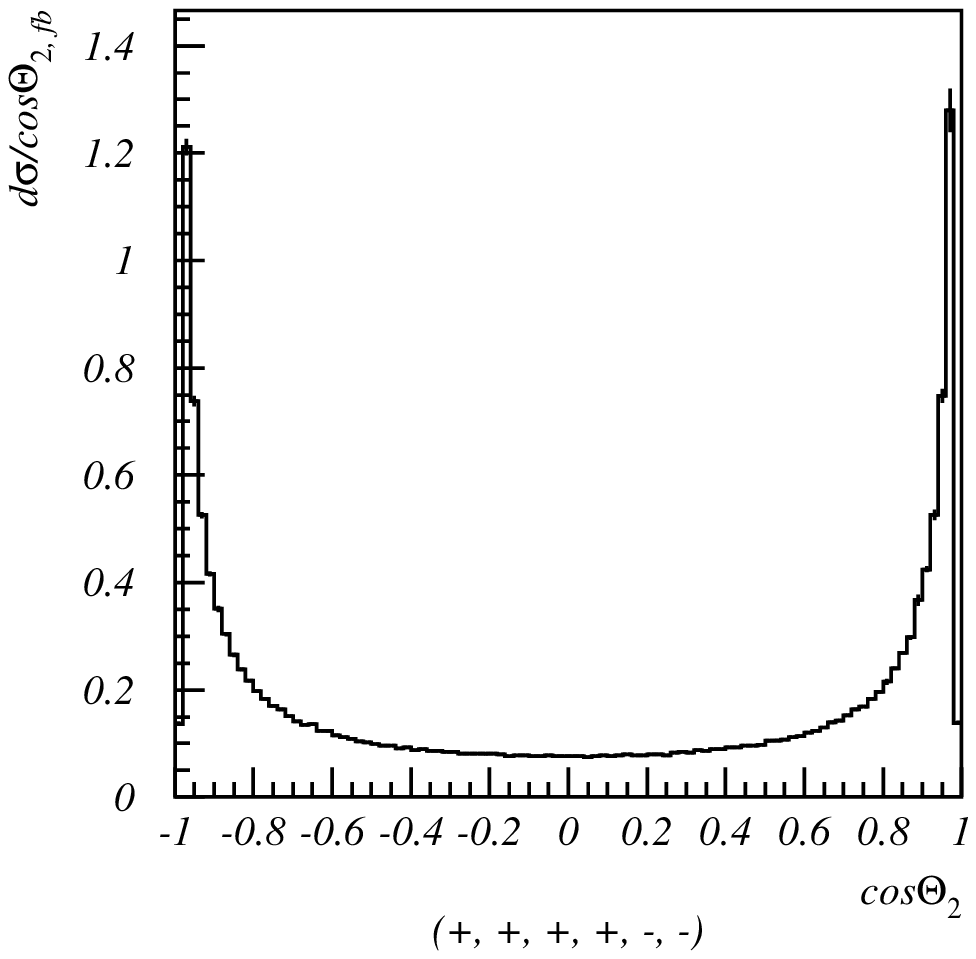}
\end{minipage}

\vspace{-10pt}
{\small Fig.4. The differential cross section of $\gamma\gamma
\rightarrow 2e^-2e^+$ process at c.m. energy of $\gamma\gamma-$
beam  $ 0.5\;{\rm TeV}$  at fixed polarization states of
interacting particles. $\theta_{1(2)}$ is the angle between the
directions of the first(second) photon and the  electron.  The
values of polar angle cut and cut of angle between any final
particles are $11^o$ and $3^o$ respectively.}
\end{figure}

It is discovered the total and differential cross sections have
strong dependence on kinematics cuts and energy of initial beam
 that table 1  clearly demonstrated.    The cross sections increase with decreasing of
energy of interacting particles because of they have reverse
dependence on scalar production $(k_1\,k_2)$  (see eq.
(\ref{c2})). Magnitude of differential cross section strongly
increases  if polar angles get close to $0$ or $\pi$ and  is on
decrease  at middle region of kinematic field (fig. 2).
Differential cross sections at fixed polarization states have
symmetric(asymmetric) form in case of similar(opposite)
polarization states of initial particles (figs. 3 and 4).
\vspace{0.5cm}

$$ $$

\end{document}